\documentstyle[prl,aps,psfig]{revtex}

\def\l{\label}

\def\vp{\varphi}

\def\lan{\langle}
\def\ran{\rangle}
\def\beq{\begin{equation}}
\def\bfg{\begin{figure}}
\def\ts{\textstyle}
\def\efg{\end{figure}}
\def\eeq{\end{equation}}
\def\bea{\begin{eqnarray}}
\def\eea{\end{eqnarray}}

\def\vphi{\langle (\delta \phi)^2 \rangle}
\def\vchi{\langle (\delta \chi)^2 \rangle}

\begin{document}
\input epsf
\draft
\vskip1pc
\title{ {\bf  Inflationary Reheating in Grand Unified 
Theories } } 
\author{
Bruce A. Bassett${}^{\P}$\footnote{email:bruce@stardust.sissa.it} 
and Fabrizio Tamburini${}^{\dag}$\footnote{email:tamburini@iol.it}
} 
\address{${}^{\P}$ International School for Advanced Studies, 
SISSA, Strada Costiera, 34014, Trieste, Italy} 
\address{${}^{\dag}$ Department of Astronomy, University of Padova, 
Padova, Italy} 
\date{\today}
\maketitle
\begin{abstract}
Grand unified theories  may display multiply    
interacting fields  with  strong coupling dynamics.  This poses 
two  new problems: (1) what is  the nature of {\em chaotic reheating} after 
inflation,  and (2) how is  reheating sensitive to  the {\em mass 
spectrum} of these theories ? We answer these questions in two 
interesting limiting  cases and demonstrate an increased efficiency of 
reheating  which  strongly enhances non-thermal  topological defect 
formation,  including monopoles and domain walls. Nevertheless,  the 
large fluctuations  may resolve this monopole problem via 
a  modified Dvali-Liu-Vachaspati mechanism in which  non-thermal  
destabilsation of discrete symmetries occurs at reheating. 
\end{abstract} 
\vspace*{1cm}
\centerline{SISSA-42/98/A}
\vspace*{1cm}
  

An ideal inflationary scenario should arise naturally from within
supergravity or a Grand Unified Theory (GUT) without fine-tuning. It 
should solve the plethora of  problems of standard cosmology while 
simultaneously diltuting the monopoles inevitably produced  due to the 
homotopic content  $\pi_2(G/U(1)) \simeq   \pi_1(U(1)) \simeq {\bf Z}$ of 
the standard model. Significant progress has been made 
within GUT's and supersymmetric theories towards  this  utopic vision 
\cite{rachel,hybrid}. However,  the issue of reheating after inflation in 
these theories,  where the universe is  revived, phoenix-like, from the 
frozen  vacuum  state, has remained  relatively unexplored 
\cite{BHP97}. This is precisely one area where the full symmetry  and 
particle content of the underlying theory is likely to be crucial. 

Indeed reheating poses a severe  threat to the simple ideal  presented 
above since non-perturbative effects are typically dominant 
\cite{KLS95,pre1,nontsr,KLS97}. Reheating is therefore {\em not} a minor 
phase at the end of inflation, of little dynamical interest.  The large 
quantum fluctuations  allow  for GUT baryogenesis \cite{KLR96,pre1} and, 
as pointed out by Kofman {\em  et al} \cite{KLS95}, may cause  the 
monopole and domain wall problems  to reappear due to non-thermal 
symmetry restoration. This last  possibility is actually rather difficult 
in simple models of reheating  with only two fields \cite{nontsr}. As we 
shall show, however, this situation changes  dramatically in the case of 
multiple fields, relevant for GUT models. 

The main motivation of this work then is to understand what new effects 
multiple fields have on reheating. This issue encompasses two particularly 
interesting unknowns. (i) The nature of reheating at 
strong coupling when the fields evolve chaotically.  This is relevant for 
GUT's  with divergent UV fixed points \cite{BOS92,BL97} and models such as 
softly broken  Seiberg-Witten  inflation \cite{GB97} where  reheating 
occurs  in the strongly coupled, confining, regime.  
Setting aside the subtle issue of the quantum behaviour of gauge 
theories at strong coupling \cite{DDR97}, in the 
two and three scalar-field cases studied  so far, (classical) chaotic 
motion  has been typical \cite{CL96}, 
especially at reheating. This  chaotic evolution parallels results in the 
Einstein-Yang-Mills  equations \cite{BL98},
semi-classical QCD  and lattice gauge theory 
\cite{qcd}. Thus a  natural question is  ``what is the 
nature of chaotic reheating ?" 

The second issue is (ii) the 
sensitivity  of reheating to the mass spectrum of the theory. For 
example, in   $SO(10)$, the one-loop  effective potential is built  from 
four quadratic invariants \cite{EMS96}. ``How does 
reheating depend on the relative masses of these four fields ?" 

We find that both these factors can significantly enhance 
the efficiency and power of reheating. This in turn improves the 
possibilities for  GUT baryogenesis and non-thermal symmetry  restoration 
(NTSR), with the  concommitant formation of topological defects  
\cite{KLS95}.

Consider the inflaton condensate $\phi$ in the
reheating phase with large couplings $\lambda_{\bf {\Psi}}$ to the 
fields $\Psi$. Further, couple $\phi$ to the minimally-coupled  scalar 
field  $\chi$ \cite{chi} via  the interaction term $\ts{1\over2}\tilde{g}^2 
\phi^2 \chi^2$, 
where $\tilde{g}$ is a dimensionless coupling constant.  Then the modes of 
$\chi$ obey \cite{BL97}:
\beq
\frac{d^2 (a^{3/2} \chi_k)}{dt^2} + \left(\frac{k^2}{a^2} + m_{\chi}^2 + 
\frac{3\kappa}{4}p + \tilde{g}^2 \phi^2 \right) (a^{3/2}\chi_k) = 0
\l{eq:quant}
\eeq
where $a(t)$ is the scale factor  of the universe obeying  the 
Hamiltonian constraint $H^2 \equiv (\dot{a}/a)^2 = 
\rho_{tot}/3$ with $\rho_{tot}$ the total energy density and 
$p$ the total pressure of the universe.  $\kappa \equiv 8\pi G = 
8\pi$. $M_{pl}$ is the Planck mass. 

In the simple model where the $\Psi$ fields  
are absent, $p = \dot{\phi}^2/2 - V(\phi)$, and if the inflaton 
effective potential is $V(\phi) =  m_{\phi}^2 \phi^2/2$, the solution is 
\cite{pre1} $\phi  \simeq \Phi(t) \sin  m_{\phi}t$,  where $\Phi$ decays 
slowly with the mean expansion.

In this case  eq. 
(\ref{eq:quant}) can be cast  in the form of the  Mathieu 
equation \cite{pre1}:  $(a^{3/2} \chi_k)'' + [A + F](a^{3/2} \chi_k) = 0$ 
with $F =  - 2q \cos(2 t')$.  The adiabatically evolving parameters 
\cite{BL97}   
$A = k^2/(m_{\phi}^2 a^2) + m^2_{\chi}/ 
m^2_{\phi} + 2\tilde{g}^2\Phi^2/m_{\phi}^2$ 
and   $q = \Phi^2(\tilde{g}^2/m_{\phi}^2 + 3\kappa/16)$,
span a plane  geometrically   dissected into  stability  
and instability  regions (fig. 1b) \cite{pre1}. The  
solutions to the Mathieu  equation in the  instability  
regions are periodic  with  envelope $y \simeq 
\exp(\mu m_{\phi} t)$, and Floquet index $\mu > 0$. This resonance 
continues  until  backreation due to $\vchi$ shuts off the resonance  by 
forcing all physical ($k \ge 0$) modes out of the dominant first resonance 
band  \cite{pre1,nontsr}.

Now switch on the couplings $\lambda_{\bf \Psi}$ to the fields 
${\bf \Psi}$ and move into the strongly coupled, chaotic region 
of the parameter space. While we are unable to study 
chaotic reheating in full generality, we have full control 
over the region in  which the chaotic fluctuations are 
extremely rapid. This is  because of the following little-known 
theorem \cite{tay}: In the limit of 
rapid variations, chaotic flows  become  uniformly 
indistinguishable  from white noise 
\cite{details}. This is an extremely  powerful 
result since the investigation of reheating becomes very pure 
and tractable.

In this limit,  eq.  (\ref{eq:quant}) again reduces to  
quasi-Mathieu form, but now with a new potential F \cite{ZMCB,bass97}: 
\beq
F = -2q \cos (2t) + g^2 \xi(t)\,.
\l{eq:def1}
\eeq
Eq. (\ref{eq:def1})  corresponds to $\lambda_{\Psi}$-independent stochastic 
evolution ($\xi(t)$  Gaussian  white noise) describing the 
effect of multiple fields in the strong coupling  limit (in 
which case $q = 0$, as there is no periodic component). This form of the 
potential also models the backreaction of quantum fluctuations on the mean 
periodic evolution of $\phi$ ($q \neq 0$) in the case where the 
$\Psi$ are absent \cite{matacz,ZMCB}.

Using spectral theory it was shown \cite{bass97} that the
stability bands essentially  disappear in the case of 
stochastic potentials $F$. In the  strong-coupling limit,  $g \rightarrow 
\infty$ we have a  rather intimidating analytical result for $\mu_k$ 
\cite{bass97} which is rather intractable and strongly emphasises the 
need for  numerical study of eq. (\ref{eq:quant}).  A selection 
from the extensive simulations performed \cite{BT98} 
for  this  problem are shown in figs.  (2a,b).  The main 
generic results are the complete breakup of the stability 
bands of the  Mathieu equations, with $\mu > 0~ \forall k$,  and the 
significant increase of $\mu$, over the purely periodic case. This is 
seen in all our simulations and is illustrated in figs. (2a) 
and (3a) which  give $\mu$ along the  physical separatrix ($k = 0$):  $A 
\simeq 2q$. 

A crucial result of these large $\mu$  is the possibility of non-thermal 
symmetry restoration (NTSR) \cite{KLS95}. Consider the simplest 2-field 
effective potential with symmetry breaking:
\beq
V(\phi,\chi) = \frac{\lambda}{4}(\phi^2 - \phi^2_0)^2 +
\frac{\tilde{g}^2}{2} \phi^2 \chi^2\,
\l{eq:pot1}\,.
\eeq
This  gains the following quantum corrections: 
$\Delta V =  3\lambda \vphi \phi^2/2 + \tilde{g}^2 \vchi \phi^2/2 + 
\tilde{g}^2  \vphi  \chi^2$/2 \cite{KLS95}, where the dominant variance is 
given in  the Hartree  approximation by \cite{KLS97} $\vphi = (2\pi^2 
a^3)^{-1}  \int dk k^2 |a^{3/2} \delta \phi_k|^2 \sim \int dk k^2 e^{4\mu 
m_{\phi} t}$.  From  energy conservation, one gets the bound  \cite{KLS95} 
$\vphi \le C \tilde{g}^{-1} \lambda M_{pl} \ln^{-2} \tilde{g}^{-2}$ with $C 
\sim 10^{-2}$. Saturating this bound  implies that Eq. (\ref{eq:pot1}) 
can have a positive effective mass, $m_{\phi, eff}^2 \equiv (V + \Delta 
V)''$  even for $\phi_0 \sim 10^{16} GeV$. 
	
Nevertheless, exhausting the available energy and realising 
these large variances in the simple Mathieu 
equations is highly non-trivial \cite{nontsr}: 
(i) sufficiently  large $\mu$ are required since $\vphi \sim 
\int dk k^2 e^{2\mu t}$. (ii)  The breadth (and very existence 
of) the first instability band controls 
how long the resonance continues before backreaction shuts it off (see 
fig 1b). (iii) The expansion of the universe is forced to be  
monotonically decreasing: $\dot{H} = - \kappa \dot{\phi}^2/2$, 
which  drives $(A,q) \rightarrow (0,0)$, damping the 
resonance and reducing the variances \cite{expand}. (iv) 
Finally, for defect production we require $m_{\phi,eff}^2 > 0$ for 
time scales $\delta t  > \omega_{\phi}^{-1}$, the period  of $\phi$ 
oscillations.

These factors conspire against NTSR and defect production in the 
simplest models of 
preheating \cite{nontsr}. However, in the strong-coupling limit of 
GUT's   considered here we know  that (i) we can easily achieve 
very large $\mu$  (see fig. 2a), (ii) since  the stability 
bands are completely destroyed 
\cite{ZMCB,bass97}, modes never stop growing, $\mu > 0$,  even 
when  backreaction becomes important. (iii) Finally, and very 
importantly, 
when there  are multiple fields, the expansion need not be 
monotonic and  indeed, in  our case will increase and decrease 
stochastically \cite{CL96} until  reheating and thermalisation 
are completed. This means that 
the expansion  need not only have a damping effect, but can also 
act as a pump, when $\dot{H} > 0$ \cite{CL96},  enhancing the 
fluctuations $\vphi$ and $\vchi$.

From these robust considerations, we expect NTSR to be much more 
effective at strong-coupling, allowing us to  approach the limiting 
variances set by energy conservation.  Indeed this is bourne out by 
direct simulation of $m_{\phi, eff}^2 \propto \lambda \vphi + \tilde{g}^2 
\vchi$  (see fig.  \ref{fig:effmstoch}). 

Examining the statistics of $m_{\phi, eff}^2$ \cite{BT98}  we 
see that while $m_{\phi, eff}^2$ fluctuates very rapidly, very large 
values of $\vchi$  occur, with  $m_{\phi,eff}^2 \in 
[-\lambda\phi_0^2, g^2 M_{pl}^2)$. Assuming $\lambda{\phi_0}^2 \ll g^2  
M_{pl}^2$,  the statistics of $m_{\phi, eff}^2$ are highly skewed, yielding  
$\overline{m_{\phi, eff}^2} > 0$. This means that most of the 
time symmetry is  restored and  $\phi$ is likely to  diffuse 
across the origin,  leading to defect production including  
monopoles in a full model including  $SU(3)_c \times  SU(2)_L  \times 
U(1)_Y$.

We now study the effect of mass spectrum deformations on 
reheating \cite{BT98}. 
A simplified toy model to 
study this issue  is given by the 3-field 
effective potential:   $V(\phi,\varphi,\chi) = 
m^2_{\phi}\phi^2/2  + m^2_{\varphi}\vp^2/2  + \tilde{g}^2 \phi^2 
\chi^2/2  + \lambda_1^2 \varphi^2 \chi^2/2$.  
Eq. (\ref{eq:quant}) is modified and leads to a new $F$ 
\cite{bass97,specdet}: 
\beq
F_{qp} = - q(\cos 2t + \cos 2\pi t)\,.
\l{eq:def2}
\eeq
where we have chosen $m_{\phi}/m_{\varphi} = \pi$, which implies that 
(\ref{eq:def2}) is a quasi-periodic function  
\cite{bass97,BT98} since the masses  are irrationally related.

In our case, 
spectral theory results guarrantee that the stable bands generically 
form a  Cantor set \cite{bass97}, which are often of very small 
measure -  hence physically  unimportant - mimicking the 
stochastic case presented earlier.  Again however, no estimates for 
$\mu$ are available analytically.  
Our numerical results show the growth of $\mu$ in this case and the 
significant widening and steepening of the instability bands relative to the 
Mathieu case (compare figs. 1 and 4). Also interesting is 
the  development of ``non-thermal edges"  in  $\mu$ (fig. 4a and 
\cite{BT98}).         

From our discussion of the stochastic case, we expect $\vphi$ to increase   
and again  this is bourne  out numerically. In figs. (5a,b) we plot $\log 
\vphi$  for the  pure Mathieu  and quasi-periodic cases as functions of 
$q$.  The maximum  variance is $\sim 1.5$  orders of magnitude larger in the 
quasi-periodic case  after the short time $t = 23 m^{-1}_{\phi}$. The 
quasi-periodic model therefore is significantly more efficient at restoring 
symmetry. In general the NTSR strength  of GUT theories will  depend 
sensitively on the mass spectrum of the 
theory, here encoded by $m_{\phi}/m_{\vp}$.

The monopole 
problem has now  again become a major concern in large (``chaotic or
incommensurate") regions of the coupling/mass parameter space, including 
the expansion of the universe.  Due to the non-thermal, quench-like, 
nature of the symmetry breaking, the 
correlation  length $\Xi$ of the fields will be much smaller than in the 
equilibrium  case, and therefore the defect density $\propto \Xi^{-n}$ 
($n = 1$ for  domain walls, $n = 3$ for monopoles) will be 
correspondingly larger than the equlibrium Kibble prediction \cite{LZ}. 

If NTSR succeeds, a 
second stage of inflation will  occur \cite{KLS95,LS95} while the vacuum 
energy $V(0)$ dominates over the energy of the $\vchi$. 
During this  time $a(t)$ increases by a factor $\sim 
(\tilde{g}^2/\lambda)^{1/4}$ for the potential (\ref{eq:pot1}), which 
cannot, therefore, supply the needed $\sim 20$ e-foldings 
to dilute the monopole density sufficiently \cite{rachel}. In 
$SO(10)$ or  $SU(6)$, however, the monopole transition is separated from 
the lower  transitions and this may allow enough secondary inflation to 
dilute the monopole  abundance sufficiently.  

However, this  is rather  model dependent and the large corrections to 
the effective  potential  offer  us an alternative  escape route via 
defect-defect interactions. The full corrections to the 
effective potential,  include not only the quadratic contributions 
affecting  $m^2_{\phi,eff}$, but  also odd powers which do not respect any 
previously existing  discrete symmetries (required for example for 
successful  D-term  inflation \cite{lyth97}). 
The odd-power corrections  to eq. (\ref{eq:pot1}), with $\tilde{g} = 0$, 
are: 
\beq
\Delta V_{odd} = \lambda \phi(\delta \phi^3 -  
\phi_0^2 \delta \phi) + \lambda \delta \phi \phi^3
\l{eq:correct2} 
\eeq
These terms softly break the ${\bf Z}_2$ $\phi \rightarrow -\phi$ 
symmetry of eq. (\ref{eq:pot1}). This allows  an 
implementation of 
the Dvali-Liu-Vachaspati mechanism \cite{DLV98} for solving the 
monopole problem as follows: imagine that   
NTSR was successful enough to produce monopoles and domain walls during 
reheating. The  $\delta \phi$  terms in eq. (\ref{eq:correct2}) 
automatically  cause the domain walls to be unstable \cite{LSW96} 
since the minima at $\pm \phi_0$ are no longer degenerate, 
causing a pressure  difference $\sim 2 \Delta 
V_{odd}(\phi_0)$ across the walls. 

Since after  preheating $\phi \ll \vphi$, the first term of 
(\ref{eq:correct2}) 
dominates.  The  constraint  that the walls percolate gives us  
$\lan\delta 
\phi\ran \le 10^{-1} M_{pl}$. Monopoles are then swept  up on  
the walls and  dissipate  because the full symmetry (e.g. 
$SU(5)$) is  restored there 
\cite{DLV98}. Requiring  that the  pressure difference 
drives the domain walls to collapse and 
decay before  dominating the energy density of the universe 
yields \cite{BT98}   
$\lan\delta \phi\ran \ge  10^{-4} \lambda^{-1/3}   M_{pl} \sim 10^{-3} 
M_{pl}$ if 
$\lambda  \sim 10^{-3}$. These bounds  are  exactly in  the range of 
values expected if NTSR is successful.  The advantage of  this  
formulation is that it  provides a specific implementation of the general 
Dvali-Liu-Vachaspati  mechanism and uses the same large quantum 
fluctuations which produced the  monopoles to remove them. 

BB would particularly like to thank Laura Covi for comments, Neil 
Cornish, Rachel Jeannerot, David Kaiser, Stefano Liberati \&  Matt Visser 
for useful discussions and George  Ellis  for hospitality at UCT.

\newpage
\begin{figure}
\epsfxsize = 5.1in
\epsffile{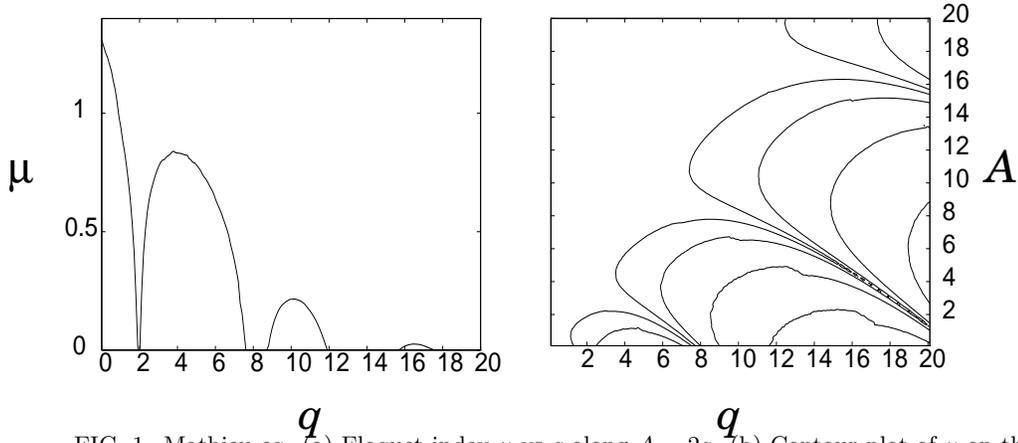}
\caption{Mathieu eq. (a) Floquet index $\mu$ vs $q$ along $A = 
2q$. (b) Contour plot of $\mu$ on the instability chart $(A,q)$.} 
\l{eq:mathspec} 
\end{figure}

\begin{figure}
\epsfxsize = 5.2in
\epsffile{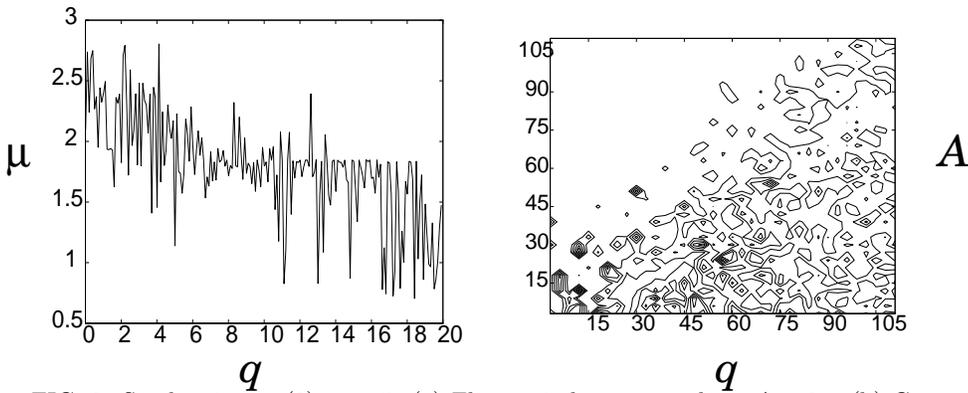}
\caption{Stochastic eq. (2), $g = 5$: (a) Floquet 
index $\mu$  vs $q$ along $A = 2q$. (b) Contour plot of 
instability chart - note the breakup of Mathieu bands and the  very large 
peaks in $\mu$.} 
\l{eq:stocspec} 
\end{figure}

\begin{figure}
\epsfxsize = 5in
\epsffile{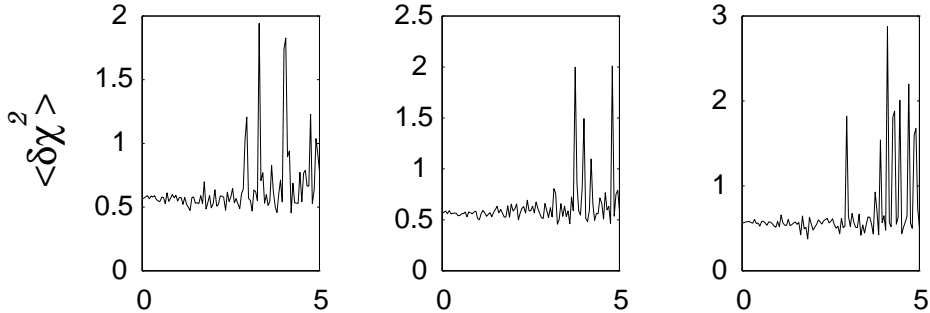}
\caption{Three  realisations of $\vchi$ vs $g$ at $t = 2m_{\phi}^{-1}$ 
for $q =5$.  At larger times,  $\vchi$ becomes completely 
dominated by a single peak (c.f. fig. 2(b)). 
} 
\l{fig:effmstoch}
\end{figure}

\begin{figure}
\epsfxsize = 5.2in
\epsffile{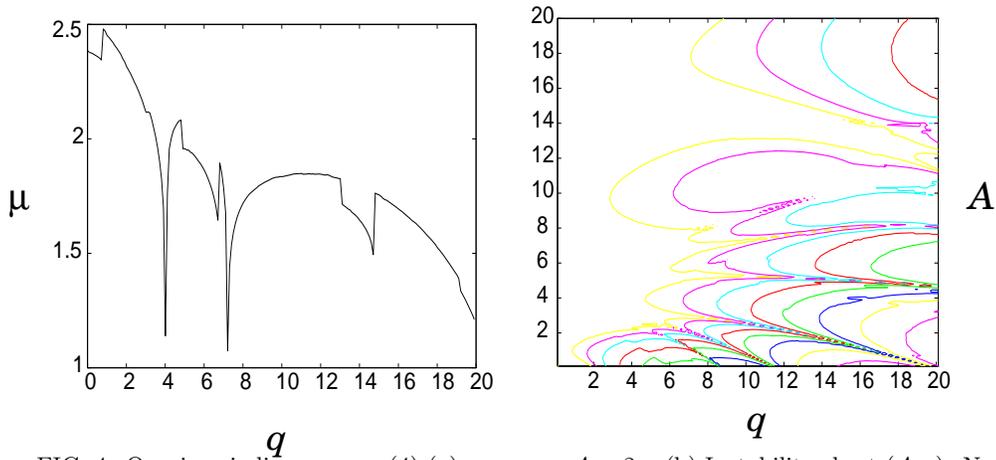}
\caption{Quasi-periodic case, eq: (4) (a) $\mu$ vs $q$ on  $A = 2q$. (b) 
Instability chart $(A,q)$. Note the larger $\mu$ and proliferation of 
instability bands compared with figs. (1 a,b).} 
\l{eq:qpspec} 
\end{figure}

\begin{figure}
\epsfxsize = 5.2in
\epsffile{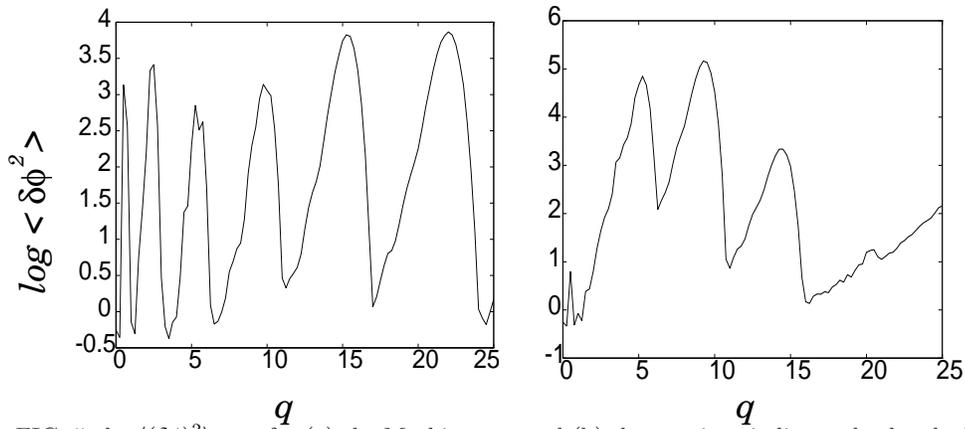}
\caption{$\log \vphi$ vs $q$  for (a) the  Mathieu eq., 
and (b) the quasi-periodic eq., both calculated  at $t = 
23m_{\phi}^{-1}$. The maximum value of $\vphi$ is about $20$ times larger 
in (b).} 
\l{eq:mathmass} 
\end{figure}


\begin{references}
\bibitem{chi} $\chi$ may be an independent field or may represent 
fluctuations $\delta \phi$ or $\delta \Psi$. $\chi$ may also be the scalar 
supersymetric partner of $\Psi$, if the $\Psi$ are fermionic.
\bibitem{details} Specifically, let $\phi$ obey Smale's Axiom 
A type  dynamics with chaotic flow $\Phi_t$, and  non-periodic  attractor 
$\Sigma$.  For an  open, dense set of functions $f$ with $E(f) = 0$, 
$\exists \sigma^2,\epsilon > 0$ and a Brownian motion $B(t)$ such that 
for almost all initial $\phi_0 \in \Sigma$:
\beq
|\int_0^t \sqrt{\lambda} f(\Phi_{\lambda\tau} \phi_0) d\tau -
\frac{\sigma}{\sqrt{\lambda}} B(\lambda t,\phi_0)| =
O(\frac{1}{\lambda^{\epsilon}})\,.
\l{eq:taylor}
\eeq
$E(f)$ denoting the expectation
value of $f$. Since white noise is the derivative of 
Brownian  motion, this implies that chaotic motion 
is, in the rapid fluctuation limit $\lambda \rightarrow \infty$, 
indestinguishable from white noise.
\bibitem{specdet} Specifically $\phi = \Phi \sin(m_{\phi} 
t)$, $\varphi = \Upsilon \sin(m_{\vp} t)$ and we neglect $p$. 
Further we choose $\tilde{g}^2  \Phi^2 = \lambda_1^2 \Upsilon^2$ to get eq. 
(\ref{eq:def2}). Relaxing this condition adds an extra parameter to eq. 
(\ref{eq:def2}) but it remains quasi-periodic, the crucial 
condition.
\end{references}
\end{document}